# A Simple Method to Check the Reliability of Annual Sunspot Number in the Historical Period 1610–1847

J. M. Vaquero[1,2] · R. M. Trigo[2,3] · M. C. Gallego[1]

[1] Departamento de Física, Universidad de Extremadura, Spain, email: jvaquero@unex.es

[2] CGUL-IDL, Universidade de Lisboa, Lisbon, Portugal

[3] Departamento de Eng. Civil da Universidade Lusófona, Lisbon, Portugal

**Abstract**

A simple method to detect inconsistencies in low annual sunspot numbers based on the relationship between these values and the annual number of active days is described. The analysis allowed for the detection of problems in the annual sunspot number series clustered in a few specific periods and unambiguous, namely: *i*) before Maunder minimum, *ii*) the year 1652 during the Maunder minimum, *iii*) the year 1741 in Solar Cycle -1, and *iv*) the so-called "lost" solar cycle in 1790s and subsequent onset of the Dalton Minimum.

## 1. Introduction

The sunspot number time series is one of the most widely studied datasets of astrophysics and geophysics specially for solar and solar-terrestrial physics. Having started in 1610, this time series is currently more than 400 years long. However it is far from being homogeneous as it was assembled by many different people, operating in different historical context and using widely different scientific tools. Clette *et al.* (2007) established four main sub-periods in the development of this time series namely: Historical period (1610–1847), Wolf period (1848–1882), Zürich period (1882–1980), and SIDC period (1981 until now). In spite of the efforts made by several scientists recovering historical observations of sunspots (*i.e.*, Hoyt and Schatten, 1998), sunspot numbers during this epoch present some problems and inaccuracies (Vaquero, 2007). Some authors have proposed different methods to allow for detection and eventual correction. In particular, Usoskin *et al.* (2003) developed a statistical method to estimate the error of the annual sunspot number even if the number of observations per year is very small. Here we proposed a different approach to detect some of these inconsistencies.

Most of these studies have used extensively the amplitudes of solar cycles on the assumption that this corresponded to the most important parameter of reconstructions of solar activity for the Historical period. However, other important characteristics of space climate are related with episodes of low or very low solar activity such as the Maunder minimum (Eddy, 1977) or the so-called "lost solar cycle" in the 1790s (Usoskin *et al.*, 2009; Zolotova and Ponyavin, 2011).

The aim of this article is to show a simple method to detect problems in low yearly sunspot numbers based in the relationship between these values and the yearly number of active days AD (days with sunspots reported on the solar disc). AD has been taken as a reliable indicator of solar activity, especially during periods of minimum activity (Maunder, 1922; Harvey and White, 1999; Usoskin, Mursula, and Kovaltsov, 2000, 2001, 2004). In fact, an equivalent index (inactive days with no spots) was used by Schwabe (1844) to discover the solar cycle.

## 2. Relationship between Sunspot Number and Active Days

The relationship between Group Sunspot Number (GSN) and active days (AD) for 1848–1995 from Hoyt and Schatten (1998) is shown in Figure 1. There is practically a linear relationship between GSN and AD when AD < 50%. Black lines in Figure 1 represent the theoretical values for an average observer with 1 (continuous), 2 (dashed), and 3 (dotted) groups for each active day. In general, it can be inferred that for years with a low AD value (*i.e.*, AD<50) we ought to expect a very low number of sunspot groups observable on the solar disc.

Note that 1848–1995 was a period of high solar activity including the recent modern Grand maximum. A similar relationship can be found if we use GSN and AD values from the Maunder minimum (1645–1715) (Figure 1, graphic inserted). Only one value (out of 71 points), corresponding to the year 1652, exceeds the lines of two (dashed line) and three (dotted line) groups per each AD. Note that some points fall slightly below the line that represents the theoretical values for an average observer detecting one sunspot group for each AD. This is due to the fact that some observers have a calibration constant $k < 1$ and there is few observers in these years.

We have applied a polynomial fit (order four) to the data of Figure 1 but restricted it to AD < 95% (blue line and corresponding points) as above 95% the relationship appears to change and break completely close to the 100%. In any case the simple relationship obtained with values AD < 95% is sufficiently strong to be used extensively in order to check yearly GSN during the historical period.

## 3. Testing Historical Values

We can draw the relationship between GSN and AD for all available annual values in the period 1610–1995 according to data provided by Hoyt and Schatten (1998) (Figure 2). To facilitate the analysis for the entire period we have attributed different colors to each century while the inset presents an amplification for the AD values smaller than 35%. It can be immediately appreciated that the entire cloud of values is spread over a larger range of values than was represented with the shorter, and better behaved 1848–1995 period in Figure 1. Moreover, it is worth noting that the 20th century values (red dots) are concentrated in the upper range of AD while the 19th values (yellow dots) are present throughout the entire regression line. On the contrary a significant amount of blue and green dots (*i.e.* 17th and 18th centuries respectively) fall unusually above or below the regression line. In fact, a number of years do not fall sufficiently close to the general relationship observed between GSN and AD described in the previous section. Among them, we highlight the years 1633, 1635, 1652, 1800, 1807 (for values of AD < 35%), and 1741, 1782, 1792, 1793 (for values of AD > 35%).

In order to quantify objectively how far from the expected values these outliers are, we have defined the anomaly ($A$) of an annual value of the GSN. We consider a year with a value $R_i$ for GSN and a value $AD_i$ for AD. Then, we can define $A = 100(R_i - R_{iFIT})/R_{iFIT}$ where $R_{iFIT}$ is the annual value of GSN obtained from the value $AD_i$ using the polynomial fit of Figure 1.

If we use this simple methodology to compute $A$, we are assuming that the annual number of days with sunspot records is not affecting this computation. However, we know that there are some years with very few observations in the historical period. Therefore, we have computed a "crude" error margin as $\Delta A = |A|/\sqrt{N}$ where $N$ is the annual number of days with sunspot records.

Anomalous values are only observed during the Historical period (1610–1848) although relatively scattered (Figure 3). We can highlight two periods with the highest concentration of anomalous values: *i*) the first years of the Maunder minimum including the previous decades leading to it and *ii*) the last decade of the eighteenth century and the first decade of the nineteenth century.

In order to further study the characteristics of these anomalous values we have computed their distribution that presents a bimodal and asymmetric shape, as shown in the graph inset in Figure 3. The vast majority of anomalies lie between -75% and 250%, however an important small secondary group of eight outliers presents outstanding anomalies between 350% and 650%, corresponding to the years 1617, 1633, 1634, 1635, 1652, 1741, 1793 and 1807. Among the few values with important negative anomalies, it is worth mentioning the year 1723 (-75%). The mean anomaly is -1.4 and the standard deviation is 18.1 for the period 1848–1995. However, if we use the complete period (1610–1995) these values will increase significantly to 47 and 110 respectively.

These outliers generally correspond to dates that may be grouped into four clusters, with different origin and time length, namely: *i*) before Maunder minimum, *ii*) the year 1652 during the Maunder minimum, *iii*) the year 1741 in Solar Cycle -1, and *iv*) the so-called "lost" solar cycle in the 1790s and subsequent onset of the Dalton minimum.

The early part of the Historical period of the sunspot record, approximately 1610–1642, is badly covered, with some years without any records at all (Hoyt and Schatten, 1998;

Vaquero, 2007). In particular, there are *i*) no records in the original GSN (Hoyt and Schatten, 1998) for the years 1636, 1637, and 1641, *ii*) a single record for the years 1610 and 1614, and *iii*) two records only for the year 1641. The average number of days with at least one sunspot record for the period 1610–1642 is 118 per year and the standard deviation is 141 records per year which represents a huge inter-annual variability. Recently, Vaquero *et al*. (2011) have made available a new sunspot number series for the period 1636–1642 using newly recovered sunspot records by Georg Marcgraf and revising sunspot data for other observation compiled by Hoyt and Schatten (1998). The Vaquero *et al*. (2011) results changed the magnitude of the sunspot cycle that occurred just before the Maunder minimum. This change indicates a gradual onset of the Maunder minimum and, thus, the reduced activity would have started two cycles before it. The problems detected in the annual value of GSN for 1617 and 1633–1635 are in agreement with this context because these years correspond to the minimum epoch of both cycles. Thus, the GSN value for these years is quite possibly overestimated and should be reduced as solar activity in both minima of the solar cycle would be fainter than what is usually taken for granted. Note that the huge reduction of the sunspot number values for the years 1638 and 1639 is not reflected in our analysis. This is due to the use by Hoyt and Schatten (1998) of estimated values (not observed values) to "adjust" the sunspot number value according to a comment by Crabtree and, therefore, AD is equal to 100% (Vaquero, 2007; Vaquero *et al*., 2011).

The simple method presented here to check the reliability of sunspot numbers indicates no relevant problems during the Maunder minimum except for the year 1652. This value corresponds to the highest anomaly for the period 1645–1700, *i.e.* the core Maunder minimum when a solar cycle is not-detectable using sunspot numbers. Again, our analysis suggests that the value for this year is most probably overestimated. Hoyt and Schatten (1998) compiled for this year sunspot records from three observers reporting zero sunspot group except Hevelius who reported five and two sunspot groups on 1 and 3 April 1652 respectively. Most probably, the values reported by Hevelius correspond to overestimates.

Some works published recently have claimed the existence of a small "lost solar cycle" in the 1790s just before the occurrence of the Dalton minimum. Usoskin *et al*. (2009) constructed the solar butterfly diagram for Solar Cycle #4 (1784–1798) using solar drawings by Staudacher and Hamilton (Arlt, 2009a, 2009b). In this diagram, the sudden occurrence of sunspots at high solar latitudes can be observed during 1793–1796. This fact was interpreted by Usoskin *et al.* (2009) as a new cycle started in 1793. However, Zolotova and Ponyavin (2011) have interpreted it as an impulse of activity in the northern hemisphere during the descending phase of this solar cycle (number 4). We have detected some anomalous values of GSN in these years and the early Dalton minimum, in particular 1793, 1807, and 1809 with anomalies above 200% (Figure 3). Therefore, according to our approach, it is highly likely that the GSN values for these years are inflated.

Another problem in the GSN series is the anomalous shape of some solar cycles and, in particular, Solar Cycle -1 (1734–1744) peaking three times, *i.e.* 1736, 1739, and 1741. Vaquero, Gallego, and Trigo (2007) localized and analyzed 1736–1739 sunspot observations published in three important scientific journals of that epoch. They obtained a new shape of Solar Cycle -1 more uniform than that in the standard GSN (Hoyt and Schatten, 1998). However, the apparent spurious peak in 1741 remained present. In this context it is particularly interesting that results obtained in this work confirm that the annual value of GSN for 1741 is anomalous and is clearly overestimated. A lower value (as provided by Usoskin, Mursula and Kovaltsov, 2004) allows for a more standard shape of Solar Cycle -1. Hoyt and Schatten (1998) compiled for this year sunspot records from M. Musano and J. Winthrop. Musano observed from 16 to 31 December (zero group sunspot for day 16 and 24–28 and one group sunspot for days 17–23 and 29–31). However, Winthrop observed seven sunspot group on 10 January 1741. This is the only record by Winthrop this year and, probably, is overestimated.

In order to provide an estimation of the more usual value of GSN for those years where anomalous values have been detected, we can assume that the values of AD from the Hoyt and Schatten (1998) database are correct and the GSN are the problematic values. In this sense, the detection or absence of sunspots on the solar disc seems a more simple information than the number of sunspot groups. Therefore, we have computed the most probable values (using the polynomial fit of Figure 1) of the eight annual values of GSN that correspond to outliers in anomaly distribution (Table 1). For the sake of comparison, we have included in Table 1 original values provided by Hoyt and Schatten (1998) and also the estimated values obtained by Usoskin *et al*. (2003).

## 4. Conclusion

Here we present a simple method to check self-consistency of annual values of GSN. Several years with unreliable GSN values have been detected. Some previous works have already drawn attention to some of these years, namely the troubled years from 1633 to 1635. Other outliers correspond to the years 1652, 1741, or 1723 (the largest negative anomaly in the series). The unusual high GSN values are almost certainly wrong and must be corrected to fall closer to the main regression line. We have computed the anomalies of these values and they surpass the values of 2 and even 3$\sigma$. Therefore, their statistical significance is greater than 99%.

Despite its simplicity and intuitiveness we must acknowledge that there are some drawbacks. This method applies only to the years of low solar activity (years with AD < 95%). Furthermore, this approach does not take into account the annual number of days of observation. In any case, all the years detected required further studies and must be addressed in future work. Here, we have presented an estimation of GSN for the problematic years. However, original sources must be consulted for a satisfactory resolution.

**Acknowledgements**

Support from the Junta de Extremadura and Ministerio de Ciencia e Innovación of the Spanish Government (AYA2008-04864/AYA and AYA2011-25945) is gratefully acknowledged.

**References**

Arlt, R.: 2009a, *Solar Phys.* **255**, 143.

Arlt, R.: 2009b, *Astron. Nachr.* **330**, 311.

Clette, F., Berghmans, D., Vanlommel, P., van der Linden, R.A.M.; Koeckelenbergh, A., Wauters, L.: 2007, *Adv. Space Res.* **40**, 919.

Eddy, J.A.: 1976, *Science* **192**, 1189.

Harvey, K.L., White, O. R.: 1999, *J. Geophys. Res.* **104**, 19759.

Hoyt, D.V., Schatten, K.H.: 1998, *Solar Phys.* **179**, 189.

Maunder, E.W.: 1922, *Mon. Not. Roy. Astron. Soc.* **82**, 534.

Schwabe, H. 1844, *Astron. Nachr.* **21**, 233.

Usoskin, I.G., Mursula, K., Kovaltsov, G.A.: 2000, *Astron. Astrophys.* **354**, L33.

Usoskin, I.G., Mursula, K., Kovaltsov, G.A.: 2001, *J. Geophys. Res.* **106**, 16039.

Usoskin, I.G., Mursula, K., Kovaltsov, G.A.: 2003, *Solar Phys.* **218**, 295.

Usoskin, I.G., Mursula, K., Kovaltsov, G.A.: 2004, *Solar Phys.* **224**, 95.

Usoskin, I.G., Mursula, K., Arlt, R., Kovaltsov, G.A.: 2009, *Astrophys. J. Lett.* **700**, L154.

Vaquero, J.M.: 2007, *Adv. Space Res.* **40**, 929.

Vaquero, J.M., Gallego, M.C., Trigo, R.M.: 2007, *Adv. Space Res.* **40**, 1895.

Vaquero, J.M., Gallego, M.C., Usoskin, I.G., Kovaltsov, G.A.: 2011, *Astrophys. J. Lett.* **731**, L24.

Zolotova, N.V., Ponyavin, D.I.: 2011, *Astrophys. J.* **736**, 115.

Table 1. Annual values of GSN for the most problematic years using original data from Hoyt and Schatten (1998) and two different estimations from Usoskin *et al.* (2004) and this work.

| Year | Hoyt and Schatten (1998) | Usoskin et al. (2004) | This work |
|------|--------------------------|------------------------|-----------|
| 1617 | 2.3  | 0.1  | 0.4  |
| 1633 | 14.3 | 0.6  | 2.3  |
| 1634 | 3.0  | 0.1  | 0.6  |
| 1635 | 4.3  | 0.1  | 0.8  |
| 1652 | 4.0  | 0.1  | 0.5  |
| 1741 | 57.7 | 10.7 | 11.9 |
| 1793 | 41.1 | 14.4 | 6.6  |
| 1807 | 5.0  | 0.2  | 0.8  |

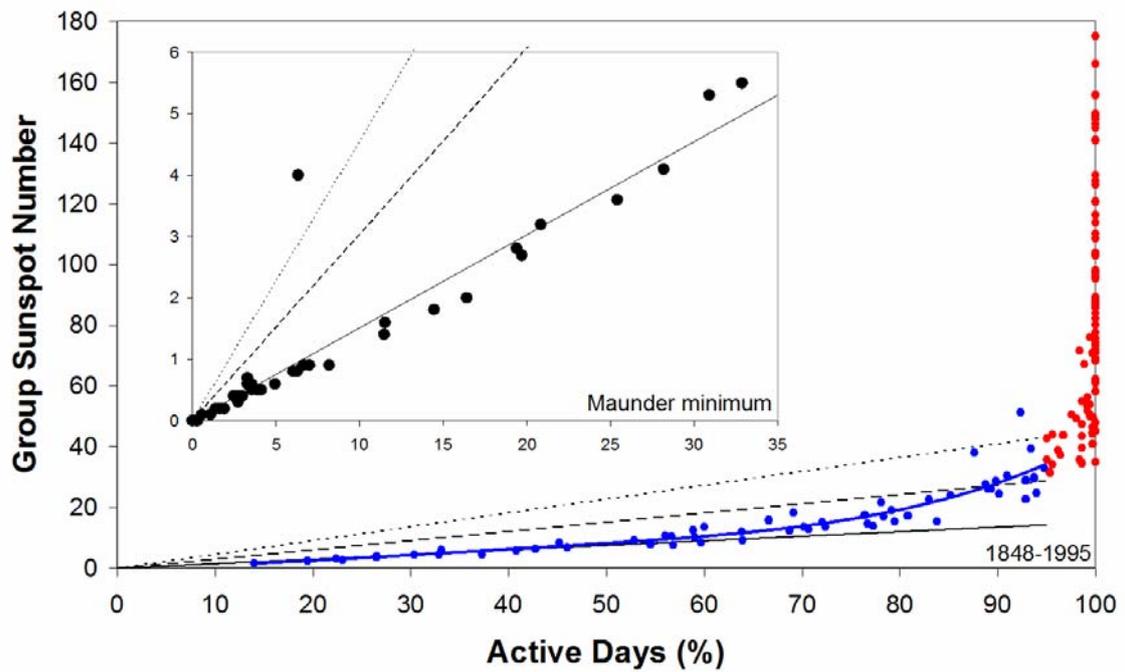

Figure 1. Relationship between GSN and AD for 1848–1995 from Hoyt and Schatten (1998). Polynomial fit (order 4) is shown for AD < 95% (blue line and points). Graphic inserted shows the same relationship during the Maunder minimum. Black lines represent the theoretical values for an average observer with 1 (continuous), 2 (dashed), and 3 (dotted) groups for each active day.

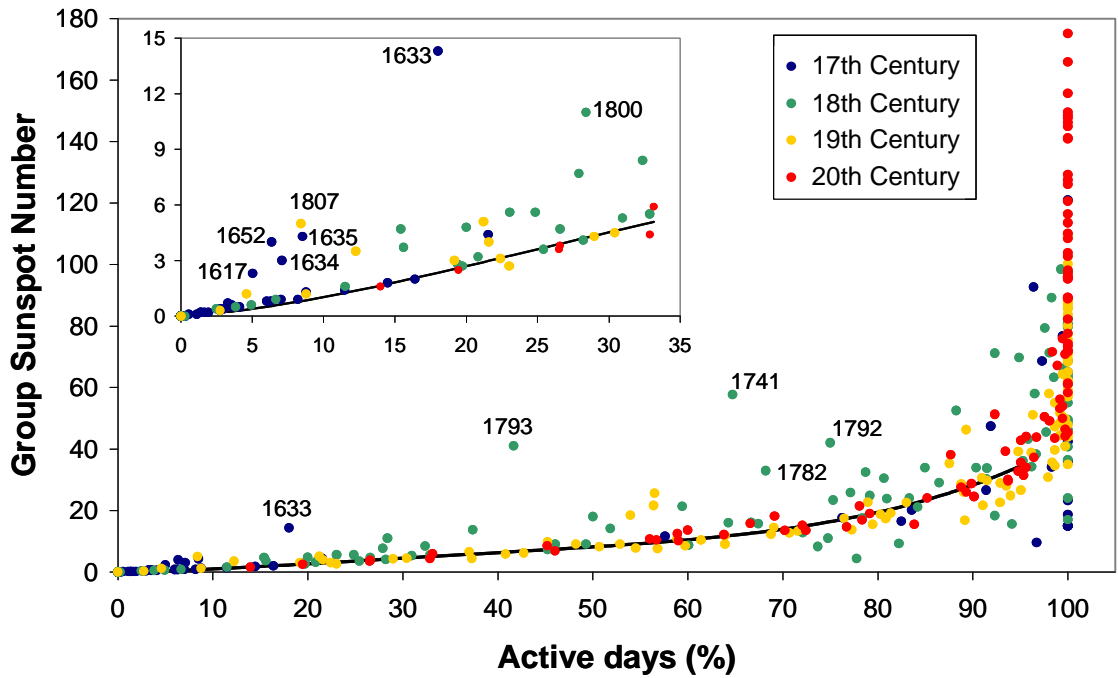

Figure 2. Relationship between GSN and AD for all available data from Hoyt and Schatten (1998). Black line is the polynomial fit of Figure 1. The inset presents an enlarged version but restricted to values AD < 35%.

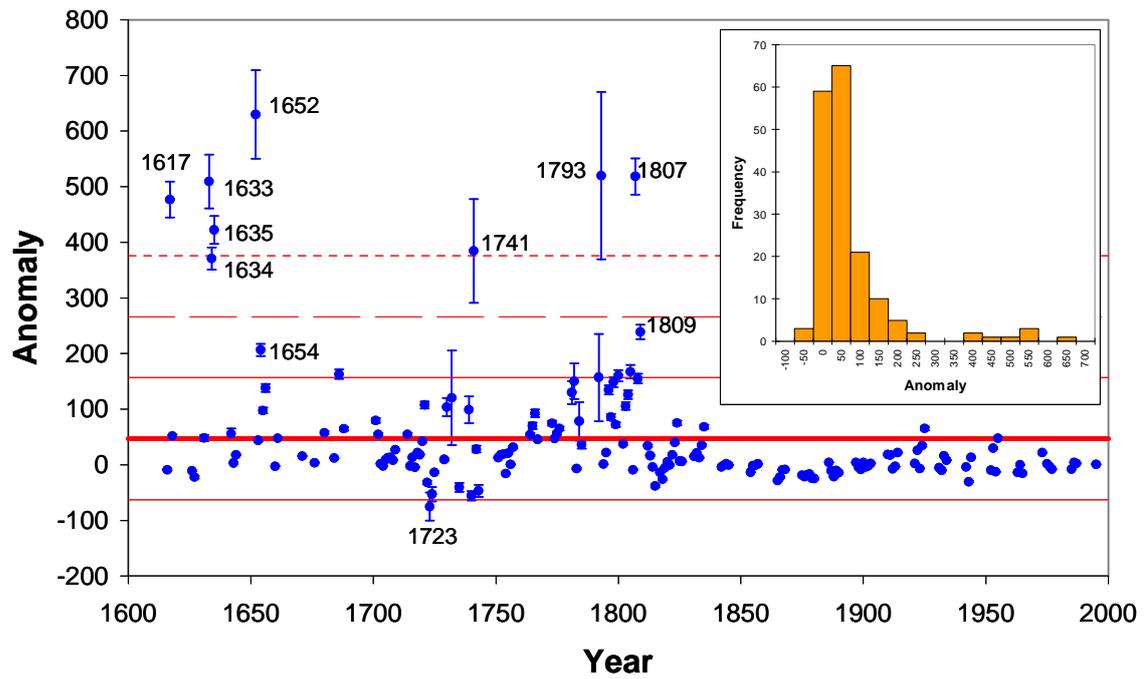

Figure 3. Anomaly values for GSN (blue dots). Insert shows the distribution of these values. Red lines represent the mean, 1σ, 2σ, and 3σ values for the complete period (1610–1995).